\documentclass[aps,reprint,superscriptaddress,nofootinbib]{revtex4-2}
\usepackage[utf8]{inputenc}
\usepackage{amsmath}
\usepackage{amssymb}
\usepackage{physics}
\usepackage{bbold}
\usepackage{graphicx}
\usepackage{multirow}
\usepackage{caption}
\usepackage{hyperref}
\usepackage{booktabs}
\usepackage{subfig}
\usepackage[export]{adjustbox}
\usepackage{todonotes}
\usepackage{tikz}
\usetikzlibrary{positioning}
\usetikzlibrary{ decorations.markings}
\usepackage{float}
\hypersetup{colorlinks=true, citecolor=blue}
\def\circledarrow#1#2#3{ 
\draw[#1,->] (#2) +(80:#3) arc(80:400:#3);
}

\newcommand{\Ua}{\uparrow}
\newcommand{\Da}{\downarrow}
\newcommand{\Ms}{{-}}

\newcommand{\SUa}{\sigma^{\Ua}}
\newcommand{\SDa}{\sigma^{\Da}}

\newcommand{\Sz}{\sigma^{z}}

\newcommand{\Sm}{\sigma^{-}}
\newcommand*{\PPS}[2][0]{%
  \foreach \superscript/\entry in {#2} {%
      \ifx\superscript\entry\sigma^{\superscript}\else\sigma^{\superscript}_\entry\fi %
  } %
}

\begin{document}

\title{Strategies for the Determination of the Running Coupling of (2+1)-dimensional QED with Quantum Computing}

\author{Giuseppe Clemente}\email{giuseppe.clemente@desy.de}
\affiliation{Deutsches Elektronen-Synchrotron (DESY), Platanenallee 6, 15738
Zeuthen, Germany}
\author{Arianna Crippa}\email{arianna.crippa@desy.de}
\affiliation{Deutsches Elektronen-Synchrotron (DESY), Platanenallee 6, 15738
Zeuthen, Germany}
\affiliation{Instit\"ut für Physik, Humboldt-Universit\"at zu Berlin,
Newtonstr. 15, 12489 Berlin, Germany}
\author{Karl Jansen}\email{karl.jansen@desy.de}
\affiliation{Deutsches Elektronen-Synchrotron (DESY), Platanenallee 6, 15738
Zeuthen, Germany}

\begin{abstract}
    We propose to utilize NISQ-era quantum devices to compute short distance quantities in $(2+1)$-dimensional QED and to combine them with large volume Monte Carlo simulations and perturbation theory. On the quantum computing side, we
perform a calculation of the mass gap in the small and intermediate regime, demonstrating, in the latter case, that it can be resolved reliably. The so obtained mass gap can be used to match corresponding results from Monte Carlo simulations, which can be used eventually to set the physical scale. In this paper we provide the setup for the quantum computation and show results for the mass gap and the plaquette expectation value. In addition, we discuss some ideas that can be applied to the computation of the running coupling.
    Since the theory is asymptotically free, it would serve as a training ground 
    for future studies of QCD in $(3+1)$-dimensions on quantum computers.
\end{abstract}

\maketitle

\section{Introduction}
In the last decades, 
non-perturbative numerical investigations 
of quantum field theories using Markov Chain Monte Carlo
(MCMC) techniques have reached unprecedented levels of 
reliability and accuracy in the determination of 
quantities of physical interest~\cite{FLAG2021}. 
On the other hand, the recent advances on 
quantum hardware open up the possibility of 
exploring new techniques and using the Hamiltonian formulation have the potential to address 
new problems such as chemical 
potentials, topological terms or real time evolution.
However, there are still many challenges that one has to face for both approaches. 
In MCMC simulations, 
autocorrelation times grow rapidly, in some cases even exponentially, (a phenomenon called \textit{critical slowing down}) 
as the lattice spacing vanishes,
making it hard to investigate the continuum limit.
Furthermore, the path integral used for 
simulating some models is afflicted by the 
infamous sign problem~\cite{PhysRevLett.94.170201,nielsen2002quantum}. 
In lattice quantum chromodynamics (LQCD),
this prevents, e.g., to accurately characterize the region of the phase diagram at finite baryon density, which is of phenomenological interest~\cite{Philipsen:2010gj,Aarts:2015tyj,Ding:2015ona}.

Regarding quantum computation, with the present hardware 
technology only very small systems can be simulated.
However, no sign problem is present 
in the Hamiltonian formalism, and no apparent obstacle 
hinders the investigation of finer lattice 
spacings thus avoiding large autocorrelation times.
For these reasons, we believe that in the current
noisy intermediate-scale quantum (NISQ) 
era~\cite{Preskill2018quantumcomputingin}
we can use quantum computing to study 
small scale properties of lattice field theories.

The system considered in this work 
is quantum electrodynamics (QED) in 2+1 dimensions,
which has been investigated in literature with different techniques~\cite{Banuls:2019bmf,PhysRevX.10.041040,PhysRevResearch.3.043209,zohar2013simulating,ambjorn1982string,loan2003path}. Our interest is in the study of small distance quantities, a most prominent example being the running coupling. By making contact with perturbation theory this allows to calculate also the $\Lambda$-parameter, $\Lambda^{QED}$ in 2+1-dimensional QED, analogous to the QCD $\Lambda$-parameter. Other possible
quantities of interest that is possible to study are, for example, the (scale dependent) renormalization factors $Z$~\cite{BOOTH2001229,PhysRevD.90.014512}.

It is the main idea of this and follow-up works, 
to match the results of short distance quantities obtained from quantum computations with the ones coming from MCMC simulations
in the strong and intermediate coupling region. In particular, when such a matching is successfully carried out, large scale MCMC calculations will provide us with a physical value of the lattice spacing. This, in turn, allows us to convert results which are obtained in lattice units, e.g. the renormalization scale or the $\Lambda$-parameter, to physical units.

The paper is organized as follows.
In Sec.~\ref{sec:system} we introduce 
the system under investigation: 
the lattice discretization
and Hamiltonian formulation are described  
in Sec.~\ref{subsec:hamiltonian},
while the electric and magnetic bases are 
discussed in Sec.~\ref{subsec:elmagbasis}.
In Sec.~\ref{sec:methods}
we introduce the numerical setup,
in particular, the encoding adopted (Sec.~\ref{subsec:encoding}) and 
the variational technique (Sec.~\ref{subsec:VQAs}). 
Numerical results are discussed in Sec.~\ref{sec:numresults},
where we show the behavior of
the spectral gap and plaquette observable 
with both exact diagonalization data and
the results of the application of variational quantum techniques on a simulator.
In Sec.~\ref{sec:outlook} 
we formulate some proposal to 
match quantum results with Monte Carlo 
simulations and to compute the running 
of the coupling in a non-perturbative way.
Finally, in Sec.~\ref{sec:conlusions}
we summarize our findings, and discuss future perspectives.

Two proposal for the improvement of the 
ansatz and for a more efficient encoding are considered in Appendix~\ref{app:gray_ansatz}  and~\ref{app:fullcompact_antipodalmagnetic} respectively,
while Appendix~\ref{app:penalty} is devoted to present tests and results for the penalty term method performed 
with the pure gauge system and the fermionic case.

\section[$(2+1)$-dimensional QED on the lattice]{(2+1)-dimensional QED on the lattice}\label{sec:system}
As for QCD in 3+1 dimensions, the behavior of QED in 2+1 dimensions 
presents confinement and asymptotic freedom. 
In this section we describe the discretization adopted~\cite{PhysRevD.90.014512}.
\subsection{Hamiltonian}\label{subsec:hamiltonian}
In order to deal with the fermionic \textit{doubling problem}~\cite{NIELSEN1981219,PhysRevD.16.3031,rothe2012lattice}, i.e. the existence in $d$-dimensions of $2^d$ flavors for each physical particle, we consider a Kogut-Susskind (K-S) formulation~\cite{PhysRevD.11.395}.
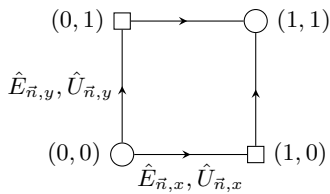
\begin{figure}[H]
    \centering
    \begin{tikzpicture}[decoration={markings, 
        mark= at position 0.5 with {\arrow{stealth}}}
    ] 
    \node[circle,draw](a0)            {};
    \node[rectangle,draw](a1)       [above =1.5cm  of a0] {} node[left =0.001cm  of a0] {$(0,0)$};
    \node[circle,draw](a2)       [right =1.5cm  of a1] {} node[left =0.001cm  of a1] {$(0,1)$};
    \node[rectangle,draw](a3)       [below =1.5cm  of a2] {} node[right=0.001cm  of a2] {$(1,1)$};
    \node(e0)       [right =0.8cm  of a3] {} node[right=0.001cm  of a3] {$(1,0)$};
    
    \draw[postaction={decorate}](a0)       -- (a1)node[midway,left]{$\hat{E}_{\vec{n}, y},\hat{U}_{\vec{n}, y}$};
    \draw[postaction={decorate}](a1)       -- (a2);
    \draw[postaction={decorate}](a3)       -- (a2);
    \draw[postaction={decorate}](a0)       -- (a3)node[midway,below]{$\hat{E}_{\vec{n}, x},\hat{U}_{\vec{n}, x}$};
    
    \end{tikzpicture}
    \caption{\textit{One plaquette system}. Fermions are on the $\vec{n}$ site, fields and the link operators $\hat{U}$ on the links. The arrows follow the positive directions $x,y$.}
    \label{fig:1plaq}
 \end{figure}
With this formulation the fermionic degrees of freedom are discretized on distinct lattices
and then separated by $2a$. Fermions and antifermions formulation are represented by single component field operators $\hat{\phi}_{\vec{n}}$, with $\vec{n}=(n_x,n_y)$ as the position on the lattice sites, as in Figure~\ref{fig:1plaq}.

The Hamiltonian that we use to represent QED~\cite[]{Haase2021resourceefficient,PRXQuantum.2.030334} consists of four terms:

\begin{equation}\label{eq:fullH}
H_{tot} = H_E + H_B + H_m + H_{kin}
\end{equation}

The first term is
\begin{eqnarray}
	\hat{H}_E = \frac{g^{2}}{2} \sum_{\vec{n}}\left(\hat{E}^{2}_{\vec{n}, x} 
	+ \hat{E}^{2}_{\vec{n}, y}\right),
\end{eqnarray}
where the field operators $\hat{E}_{\vec{n}, \mu}$ (direction $\mu=x,y$) are dimensionless and act on the links, and $g /\sqrt{a}\to g$ is the coupling constant. The electric fields take integer eigenvalues $e_{\vec{n}}=0,\pm 1,\pm 2,...$ with $\hat{E}_{\vec{n}, {\mu}}\ket{e_{\vec{n}}}=e_{\vec{n}}\ket{e_{\vec{n}}}$.

In LGTs, the following Wilson operators are introduced~\cite{PhysRevD.10.2445} on the links,
\begin{eqnarray}
\hat{U} = \mathrm{e}^{iag\hat{A}_{\mu}(\vec{n})},
\end{eqnarray}
where $\hat{A}_{\mu}(\vec{n})$ is the vector field. 
They measure the phase proportional to $g$ acquired by a unit charge moved along the link and act as a lowering operator on electric field eigenstates: $\hat{U}_{\vec{n}, {\mu}}\ket{e_{\vec{n}}}=\ket{e_{\vec{n}}-1}$. Thus, we have the following commutator
\begin{equation}
    [\hat{E}_{\vec{n}, \mu},\hat{U}_{\vec{n}^\prime, \nu}] = - \delta_{\vec{n},\vec{n}^\prime} \delta_{\mu,\nu}\hat{U}_{\vec{n}, \mu}.
\end{equation}

The magnetic interaction is determined by the plaquette operator $\hat{P}_{\vec{n}} =  \hat{U}_{\vec{n},x}\hat{U}_{\vec{n}+x,y}\hat{U}^{\dag}_{\vec{n}+y,x}\hat{U}^{\dag}_{\vec{n},y}$,
\begin{eqnarray}
\hat{H}_B = -\frac{1}{2g^{2}} \sum_{\vec{n}} \left(\hat{P}_{\vec{n}} + \hat{P}_{\vec{n}}^{\dag}
\right).
\end{eqnarray}

For the fermionic part of the Hamiltonian we have the mass term 
\begin{eqnarray}
\hat{H}_{m} = m \sum_{\vec{n}} (-1)^{n_x+n_y} \hat{\phi}^\dag_{\vec{n}} \hat{\phi}_{\vec{n}},
\end{eqnarray}
and the kinetic term,
\begin{eqnarray}
\hat{H}_{kin} = \Omega \sum_{\vec{n}} \sum_{\mu = x,y} (\hat{\phi}_{\vec{n}}^{\dag} \hat{U}_{\vec{n},
 \mu} \hat{\phi}_{\vec{n}+ \mu} + {H.c.}),
\end{eqnarray}
where $[\Omega] = [1/(2a)]$. 
It corresponds to the creation or annihilation of a fermion-antifermion pair on neighbouring lattice sites and the adjustment of the link.

\vspace{0.4cm}
The Hamiltonian is gauge invariant, i.e. it commutes with the Gauss' law operators $G_{\vec{n}}$ at
each site $\vec{n}$
\begin{equation}\label{eq:Gauss}
\begin{aligned}
 G_{\vec{n}}=\Bigg[\sum\limits_{\mu=x,y}&
\left(\hat{E}_{\vec{n}, \mu} -\hat{E}_{\vec{n}- \mu, \mu} \right) - \hat{q}_{\vec{n}} - \hat{Q}_{\vec{n}}\Bigg] \\
& G_{\vec{n}}\ket{\Phi} = 0 \iff \ket{\Phi} \in \mathcal{H}_{\text{phys.}},
\end{aligned}
\end{equation}
where $\hat{q}_{\vec{n}}$ are \textit{dynamical charges}, defined as 
\begin{equation}\label{eq:dyn_charges}
    q_{\vec{n}} = \hat{\phi}^\dag_{\vec{n}} \hat{\phi}_{\vec{n}} - \frac{1}{2}(1+(-1)^{\lvert \vec{n} \rvert +1}).
\end{equation}
$\hat{Q}_{\vec{n}}$ are external
\textit{static charges}, 
which are set to non-zero values only 
when one considers 
the computation of the static potential.

\subsection{Electric and magnetic basis and fermion discretization}\label{subsec:elmagbasis}

In this section we consider the calculation of the energy gap with periodic boundary conditions (PBC). 
These allow, unlike open boundary conditions, 
to properly define the momentum operator and hence 
project into 
the zero-momentum sector ($\vec{p}_{\text{tot.}}=\vec{0}$) along with 
the total zero-charge sector ($Q_{\text{tot.}}=0$).
In principle it should be easier to extract the energy difference when the number of states to be considered is smaller. 
However, this procedure is not straightforward in the open boundary case, since no momentum operator in the sense of generator of translations exists; in this case, 
it is still possible to introduce a pseudo-momentum operator~\cite{Ba_uls_2013} 
whose square can be used to select 
only the physical states among all the results
a posteriori, or as a suppression term
with the same procedure described 
in Sec.~\ref{subsec:penalty}.

\begin{figure}[H]
    \centering
    \begin{tikzpicture}[decoration={markings, 
        mark= at position 0.5 with {\arrow{stealth}}}
    ] 
    
    \node[circle,draw](a0)            {};
    \node[rectangle,draw](a1)       [above =1.5cm  of a0] {};
    \node[circle,draw](a2)       [right =1.5cm  of a1] {};
    \node[rectangle,draw](a3)       [below =1.5cm  of a2] {};
    \node(a5)       [right =1.5cm  of a2] {};
    \node(a4)       [right =1.5cm  of a3] {};
    \node(a6)       [above =1.5cm  of a1] {};
    \node(a7)       [above=1.5cm  of a2] {};
    \node(a8)       [right=1.5cm  of a2] {};
    \node(i1)       [above right=0.7cm  of a0] {\textcolor{red}{$R1$}};
    \node(i2)       [above right=0.7cm  of a3] {\textcolor{red}{$R2$}};
    \node(i3)       [above right=0.7cm  of a2] {\textcolor{red}{$R3$}};
    \node(i4)       [above right=0.7cm  of a1] {\textcolor{red}{$R4$}};
    
    \node(s1)       [left=0.1cm of a0] {};
    \node(s2)       [left=0.1cm of a6] {};
    \node(s3)       [below=0.1cm of a0] {};
    \node(s4)       [below=0.1cm of a4] {};
    
    \circledarrow{ultra thick, red}{i1}{0.6cm};
    \circledarrow{ultra thick, red}{i2}{0.6cm};
    \circledarrow{ultra thick, red}{i3}{0.6cm};
    \circledarrow{thick, red}{i4}{0.6cm};
    
    \draw[postaction={decorate}](a0)       -- (a1);
    \draw[postaction={decorate}](a1)       -- (a2);
    \draw[postaction={decorate}](a3)       -- (a2);
    \draw[postaction={decorate}](a0)       -- (a3);
    \draw[postaction={decorate}](a3)       -- (a4);
    \draw[postaction={decorate}](a2)       -- (a5);
    \draw[postaction={decorate}](a1)       -- (a6);
    \draw[postaction={decorate}](a2)       -- (a7);
    
    \draw[ultra thick,red,postaction={decorate}](s1)       -- (s2)node[midway,left]{$R_y$};
    \draw[ultra thick,red,postaction={decorate}](s3)       -- (s4)node[midway,below]{$R_x$};

    \end{tikzpicture}
    \caption{\textit{$2 \times 2$ PBC lattice}. By applying the Gauss' law we can write $R4$ (thin line) in terms of the other three rotators.}
    \label{fig:PBCplaq}
 \end{figure}
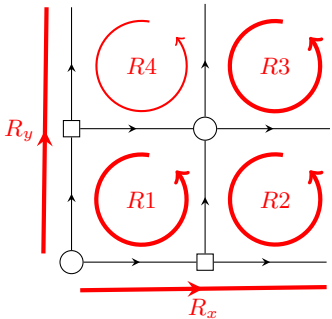

\vspace{0.4cm}
We consider a square lattice formed by four lattice sites. Due to Gauss' law, only five of the eight links are independent, thus we can have a more resource efficient Hamiltonian, suitable for a wide range of quantum hardware. To conveniently simplify the expressions we use the set of operators, i.e. \textit{rotators} and \textit{strings}~\cite{Haase2021resourceefficient}, which preserve Gauss' law, as in Figure~\ref{fig:PBCplaq}. 
Using periodic boundary conditions, the 
momentum-zero sector can be enforced exactly 
by using a translational invariant Hamiltonian\footnote{
However, after gauge fixing and for a finite truncation, symmetry cannot be enforced exactly on the remaining gauge degrees of freedom (i.e., rotators and strings).}.

As mentioned, our analysis include fermionic matter and excited states and by using a K-S formulation we have a symmetric fermionic system every two lattice spacings.

\vspace{0.4cm}
\subsubsection{Discretization and Truncation}\label{subsubsec:truncation}

In the following, we describe two schemes that allow to discretize the continuous $U(1)$ group with $\mathbb{Z}_{2L+1}$ which provides a discrete basis for the operators $\hat{A}_{\vec{n}, {\mu}}$. 
Since the gauge rotators possess discrete but infinite
spectra, any numerical approach requires a truncation of the Hilbert space. Thus the gauge fields assume integer values in range $[-l,l]$, and the dimension of the Hilbert space (within the gauge part) is $(2l+1)^5$. The parameter $l$ is the cut-off value for the group truncation. 

We consider two different representations for the Hamiltonian, i.e. electric and magnetic. While in the former the operators associated to the electric field are diagonal, the plaquette operator corresponds to a non-diagonal expression. As a consequence, applying a different representation in the weak coupling regime is preferable. This is performed through the discrete Fourier transform, that diagonalizes the lowering operators. 

The discrete approximation in the magnetic basis is related to the parameters $l$ and $L$.
We refer to $l$ as the truncation level and $L$
as the discretization level.
Since the truncated $U(1)$ and $\mathbb{Z}_{2L+1}$ are equivalent in the electric basis, the parameter $L$ is irrelevant, however it strongly influences the results derived in the magnetic representation. 
For a given $l$, $L$ defines the resolution of the approximation centered around the vacuum. The group is discretized into $2L+1$ states and only $2l+1$ are kept after the truncation.

\subsubsection{Jordan-Wigner}

For the numerical implementation, the fermionic degrees of freedom can be mapped to spins using a Jordan-Wigner transformation~\cite{jordan1993paulische},
\begin{align}\label{eq:jw}
    \phi_{\vec{n}}          &= \Big[\prod\limits_{{\vec{k}} <{\vec{n}} }( i\sigma^z_{\vec{k}} )\Big] \sigma^{-}_{\vec{n}} ,\\
    \phi_{\vec{n}}^\dagger  &= \Big[\prod\limits_{{\vec{k}} <{\vec{n}} }(-i\sigma^z_{\vec{k}} )\Big] \sigma^{+}_{\vec{n}} ,
\end{align}
where $\sigma^{z}$ is the $z-$Pauli matrix, $\sigma^{\pm}\equiv \frac{\sigma^{x}\pm i\sigma^{y}}{2}$ and the relation $\vec{k} < \vec{n}$ is defined by $(0, 0) < (0, 1) < (1, 1) < (1, 0)$ to satisfy the fermionic commutation relations.

Applying Eq.~\eqref{eq:jw} to~\eqref{eq:dyn_charges},
the expressions for the dynamical charges 
in terms of bosonic spin operators become:
\begin{equation}
    q_{\vec{n}} = \frac{1}{2}(\sigma_{\vec{n}}^z+(-1)^{\lvert \vec{n} \rvert +1}).
\end{equation}

\section{Numerical setups and methods}\label{sec:methods}
\subsection{Encoding}\label{subsec:encoding}

The encoding adopted in~\cite{PRXQuantum.2.030334} maps the $N$ fermionic states into an equal number of qubits and gauge physical states onto $2l+1$ qubits using
\begin{align} \label{eq:encodedstates}
  \ket{-l + j}_{\text{phys.}} &\mapsto \ket*{\overbrace{0 \hdots 0}^{j} 1 \overbrace{0 \hdots 0}^{2l-j}}.
\end{align}
The truncated electric field and link operators (for a single edge) can be expressed starting from their expressions in the physical basis:
\begin{align}
    \hat{E} &= \sum_{i=-l}^l i \ket{i}_{\text{phys.}}\bra{i}_{\text{phys.}}, \\
    \label{eq:U_electric}
    \hat{U} &= \sum_{i=-l+1}^l \ket{i-1}_{\text{phys.}}\bra{i}_{\text{phys.}}.
\end{align}
For example, for $l=1$ the corresponding $2l+1$ possible states are encoded as
\begin{align}
    \ket{-1}_{\text{phys.}} \mapsto &~\ket{100}, \\
    \ket{0}_{\text{phys.}} \mapsto &~\ket{010}, \\
    \ket{1}_{\text{phys.}} \mapsto &~\ket{001}. 
\end{align}
However, this mapping is not resource efficient since we would need $2l+1$ qubits for each gauge variable.

Hereby we use another method to encode the physical states. 
Instead of using $2l+1$ qubits for each truncated gauge variable, 
one would like to use a compact encoding, see e.g.~\cite{PhysRevA.103.042405}.
The minimum number of qubits required per gauge variable is\footnote{In principle, one could even encode in the qubit space any number $N$ 
    of gauge variables, so that the space one uses $\lceil \log_2[N(2l+1)] \rceil$ qubits 
instead of $N \lceil \log_2(2l+1) \rceil$, gaining some additional space, but not much.}
$q_\text{min} = \lceil \log_2(2l+1) \rceil$,
such that, using $N$ truncated gauge variables, 
the fraction of physical states ${(2l+1)}^N$ among the total number
of states in the encoding is $2^{N(\{\log_2(2l+1)\}-1)}$, 
which corresponds to $\frac{1}{2^N}{(1-\frac{1}{4l})}^N$ 
in the worst case (when $l$ is a power of $2$); using the previous encoding instead, 
one would use just a fraction $\frac{{(2l+1)}^N}{2^{N (2l+1)}}$, which vanishes exponentially 
in the limit $l \to \infty$ with a rate coefficient $2N\ln 2$.

In order to represent a generic transition
in the following discussion, we use the notation
\begin{equation}\label{eq:sigma_defs}
    \begin{aligned}
    \sigma^{+} &\equiv \ketbra{1}{0},\, 
    & \SUa       &\equiv \ketbra{1}{1},\, \\
    \sigma^{-} &\equiv \ketbra{0}{1},\, 
    & \SDa       &\equiv \ketbra{0}{0}.
    \end{aligned}
\end{equation}
Amongst all the possible 
$\binom{2^{q_\text{min}}}{2l+1}$ 
compact encodings, it is convenient to pick 
the ones that simplify the ladder terms 
$\ket{i-1}_{\text{phys.}}\bra{i}_{\text{phys.}}$ in 
Eq.~\eqref{eq:U_electric} 
in such a way that the bit string representing
the state in the computational basis changes only
by one bit. Another way of phrasing this condition 
is that we would minimize  the number of 
off-diagonal operators appearing
in the Pauli string decompositions of these terms.
This is realized by the so called \textit{Gray codes}.

For example, with $l=1$ there are 
three physical states  $\ket{i}_{\text{phys.}}$ 
for $i\in \{-1,0,1\}$, 
which can be encoded using only $2$ qubits using a Gray encoding pattern as shown in Table~\ref{tab:gray_enc1},
where the state $\ket{10}$ is considered unphysical.
\begin{table}[ht]
    \centering
    \caption{Gray encoding for $l=1$.}\label{tab:gray_enc1}
    \begin{tabular}{c | c | c | c}
        \toprule
        $\ket{i}_{\text{phys.}}$ & $\ket{i}$ & 
        $\ketbra{i}$ & $\ketbra{i-1}{i}$\\
        \midrule
        $\ket{-1}_{\text{phys.}}$ &  $\ket{00}$     & $\sigma^\Da \otimes \sigma^\Da$  &  $0$ \\
        $\ket{0}_{\text{phys.}}$  &  $\ket{01}$     & $\sigma^\Da \otimes \sigma^\Ua$  &  $\sigma^\Da \otimes \sigma^\Ms$ \\
        $\ket{+1}_{\text{phys.}}$ &  $\ket{11}$     & $\sigma^\Ua \otimes \sigma^\Ua$  &  $\sigma^\Ms \otimes \sigma^\Ua$ \\
        unphysical &  $\ket{10}$   &   &  \\
        \bottomrule
    \end{tabular}
\end{table}

The expressions for the truncated electric field and link operators then become:
\begin{align}
    \hat{E} &\mapsto -\ketbra{00}{00} +\ketbra{11}{11}
    = -\frac{1}{2} \Big[\Sz_0 + \Sz_1\Big], \\
    \hat{U} &\mapsto \ketbra{00}{01} +\ketbra{01}{11} =\frac{1}{2} 
    \Big[\Sm_0(I_1 + \Sz_1)+ \Sm_1 (I_0 - \Sz_0) \Big].
\end{align}
Unfortunately, since $2l+1$, the number 
physical basis states for gauge variables, is not 
a power of two, one has to deal with the unphysical
states so that they would not appear in the final solutions. 
This can be done either by working with a structured ansatz with rotations allowed only between physical states (see Appendix~\ref{app:gray_ansatz})
or by introducing a penalty term in the Hamiltonian which suppress solutions with overlap in the unphysical region of the Hilbert space (see Sec.~\ref{subsec:penalty}).
Another approach which can be applied in the case 
of magnetic basis consists in changing the discretization in such a way that one can 
include the remaining $2^{\lceil \log_2(2l+1) \rceil} -2l-1$ states as
physical (see Appendix~\ref{app:fullcompact_antipodalmagnetic}).

\subsection{Variational Quantum Algorithms}\label{subsec:VQAs}
In this section we describe the technique adopted to 
compute the mass gap and the expectation value 
of the plaquette operator, using an extension 
of the Variational Quantum Eigensolver (VQE) 
algorithm~\cite{Peruzzo_2014}.
A quantum device can be used to efficiently evaluate the expectation value of tensor products of an arbitrary sequence of Pauli operators~\cite{ortiz2001quantum}. Since the Hamiltonian is written as a weighted sum of such Pauli string terms, also $\langle H \rangle$ can be efficiently estimated\footnote{In general, it is possible to choose a \textit{grouping strategy} to identify subsets of Pauli strings appearing in the Hamiltonian, it is possible to reduce the number of independent circuit evaluations~\cite{vqereview}.}. 

In order to find the lowest eigenvalue of a given operator $H$, the variational approach finds an approximation to the eigenvector $\ket{\psi}$ which corresponds to the lowest eigenvalue and that minimizes 
\begin{equation}
    E(\theta):=\expval{H}{\psi(\theta)},
\end{equation}
where the state
$\ket{\psi({\theta})}=U(\theta)\ket{0}$ 
is realized as a parameterized circuit $U(\theta)$ called \textit{ansatz}.

This procedure is done by varying a vector $\theta$ of scalar parameters (typically gate rotation angles) through the combination of a classical and a quantum part.
 
The Variational Quantum Deflation 
(VQD) method~\cite{PhysRevA.99.062304,Higgott2019variationalquantum} 
extends VQE to estimate the $k$-th excited state $E_k$ by penalizing the solutions of the lowest excited states. This is done through a minimization of the cost function
\begin{equation}
    C(\theta_k)=\expval{H}{\psi(\theta_k)} + \sum_{i=0}^{k-1}\beta_i {\big\lvert \braket{\psi(\theta_k)}{\psi(\theta_i^*)}\big\rvert}^2,
\end{equation}
where $\beta_i$ are real-valued coefficients (which must be larger than the gaps $E_k-E_i$) and $\theta_i^*$ are the optimal parameters for the $i$-th excited state. The overlap terms are computed by either using the inverse circuit\footnote{Here and in our results we assume the parameterized ansatz to be the same $U(\theta)$ for every excited state, but this requirement is not necessary.} $U^\dag(\theta_k) U(\theta_i^*) \ket{0}$ and estimating the occurrence of all-zero measurements or by a SWAP test~\cite{Higgott2019variationalquantum,Havl_ek_2019,PhysRevA.75.012328}.
This can be interpreted as minimizing the $E(\theta_k)$ with the constraint that $\ket{\psi(\theta_k)}$ must be orthogonal to the previous $k$ states. 
Since our goal is to compute the energy gap between the ground state $E_0$ and first excited state $E_1$, we follow three main steps:
\begin{itemize}
\item[1.] Perform the VQE and obtain optimal parameters and an approximate ground state 
$\ket{\psi(\theta_0^*)}$;
\item[2.] For $E_1$ define a Hamiltonian:
\begin{equation}
    H_1 = H + \beta \ket{\psi(\theta_0^*)} \bra{\psi(\theta_0^*)};
\end{equation}
\item[3.] Perform the VQE with 
the Hamiltonian $H_1$ to find an approximation of the first excited state $\bra{\psi(\theta_1^*)}$.
\end{itemize}
In our case, for each value of the coupling $g$ we can compute the best approximation to the ground state and extract
$E_0$ and the expectation value of the plaquette $\expval{\Box}$,
while from the first excited state we
just need $E_1$ to estimate the spectral 
gap $\Delta E = E_1 - E_0$.

\subsection{Ansatz and penalty term}\label{subsec:penalty}
In this section we describe the procedure 
that we used to deal with unphysical states.
Instead of constraining the reachable states to the 
physical ones at the level of
the ansatz (as discussed in Appendix~\ref{app:gray_ansatz} for the gray encoding),
we consider a generic ansatz and introduce,
directly in the definition of the Hamiltonian,
a penalty term that suppresses unphysical 
contributions on the final states~\cite{PhysRevResearch.3.043209}.
The form that we used for this suppression term is the following:
\begin{equation*}
          \Delta H_{\text{suppr.}} = \lambda \Big[ \sum\limits_{p\in {\text{gauge}}} \Pi^{(u)}_{p} +   \Pi^{(Q_{\text{tot}}\neq 0)}_{\text{ferm.}}\Big],
      \end{equation*}
      where $\lambda$ is the \textit{suppression coefficient},
      while $\Pi^{(u)}_{p}$ and $\Pi^{(Q_{\text{tot}}\neq 0)}_{\text{ferm.}}$ are respectively the projectors onto the unphysical Hilbert spaces of single gauge variables (identity on the other variables) and the projector onto the non-zero charge space.

At the end of the VQE optimization,
we can assess how much the optimal state reached $\ket{\psi(\theta^*)}$
is \textit{unphysical} 
by measuring its overlap with the unphysical Hilbert space $\mathcal{H}_{\text{unphys.}}$. 
In practice, this is done by computing the 
expectation value
of the projector 
into $\mathcal{H}_{\text{unphys.}}$,
\begin{equation}
u(\theta^*_k) \equiv \expval{\Pi_{\text{unphys.}}}{\psi(\theta^*_k)}.
\end{equation}

First, we tested this method for the pure gauge system, then we considered the fermionic case with a selection procedure for $\lambda$ based on a bisection approach. See Appendix~\ref{app:penalty} for details and results.

\section{Numerical results}\label{sec:numresults}
Here we discuss results of exact 
diagonalization (ED) of the Hamiltonian Eq.~\eqref{eq:fullH} in both electric and 
magnetic basis and with $\Omega = 1$, $m = 0$. 
Then we show some results of the VQD method.

\subsection{Exact diagonalization results}\label{subsec:ed}
\begin{figure}[ht]
    \centering
    \includegraphics[width=1\linewidth]{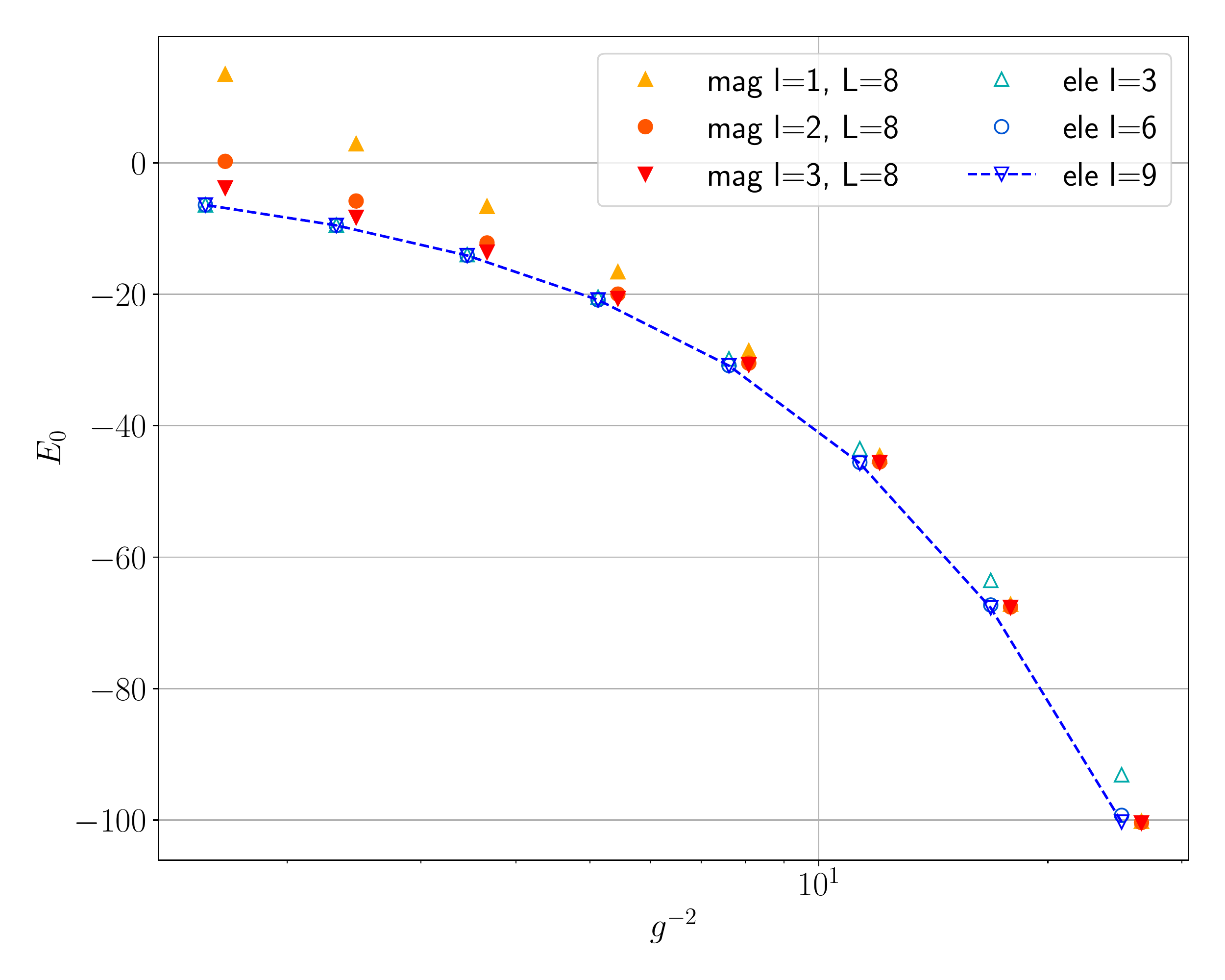}
    \caption{Exact diagonalization data for ground state energy in both electric and magnetic basis and selected values of discretization and truncation (see legend).}%
    \label{fig:plot_match_ele_mag_QED_PBC_2x2_E0}
\end{figure}

\begin{figure}[ht]
    \centering
    \includegraphics[width=1\linewidth]{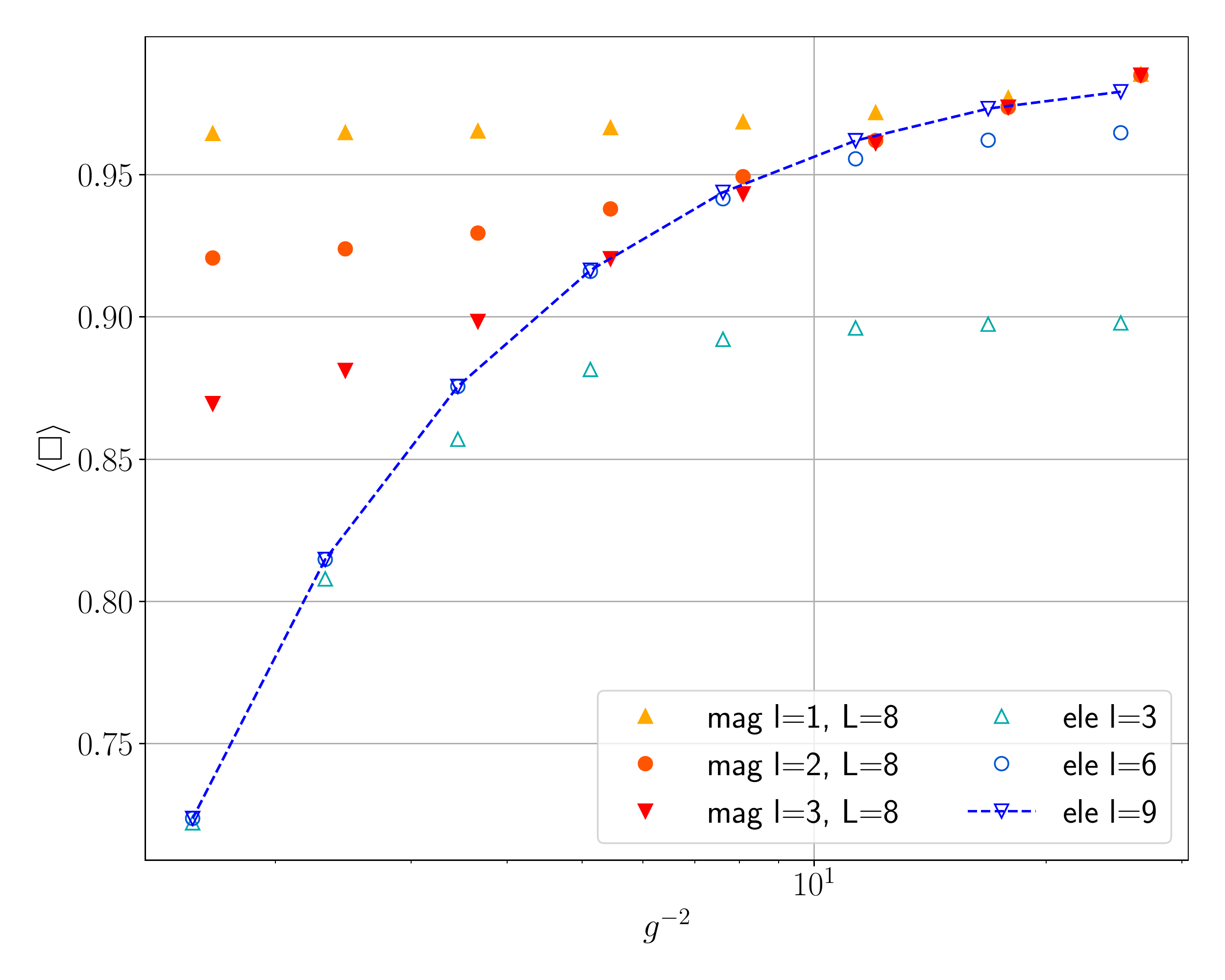}
    \caption{Exact diagonalization data for the plaquette in both electric and magnetic basis and selected values of discretization and truncation (see legend).}%
    \label{fig:plot_match_ele_mag_QED_PBC_2x2_Pl}
\end{figure}

Figure~\ref{fig:plot_match_ele_mag_QED_PBC_2x2_E0}
and Figure~\ref{fig:plot_match_ele_mag_QED_PBC_2x2_Pl} 
show respectively the ground state energy
and plaquette expectation value
for different levels of truncation $l$,
and for the discretization level $L=8$ in the magnetic basis.
As expected, for small values of the
coupling $g$ the convergence in the truncation parameter $l$ at fixed discretization $L$ is 
generally faster in the magnetic basis. In the case 
of  larger couplings the electric basis 
performs better, 
due to the relative dominance of the
$H_E$ or $H_B$ terms in the Hamiltonian.
\begin{figure}[ht]
    \centering
    \includegraphics[width=1\linewidth]{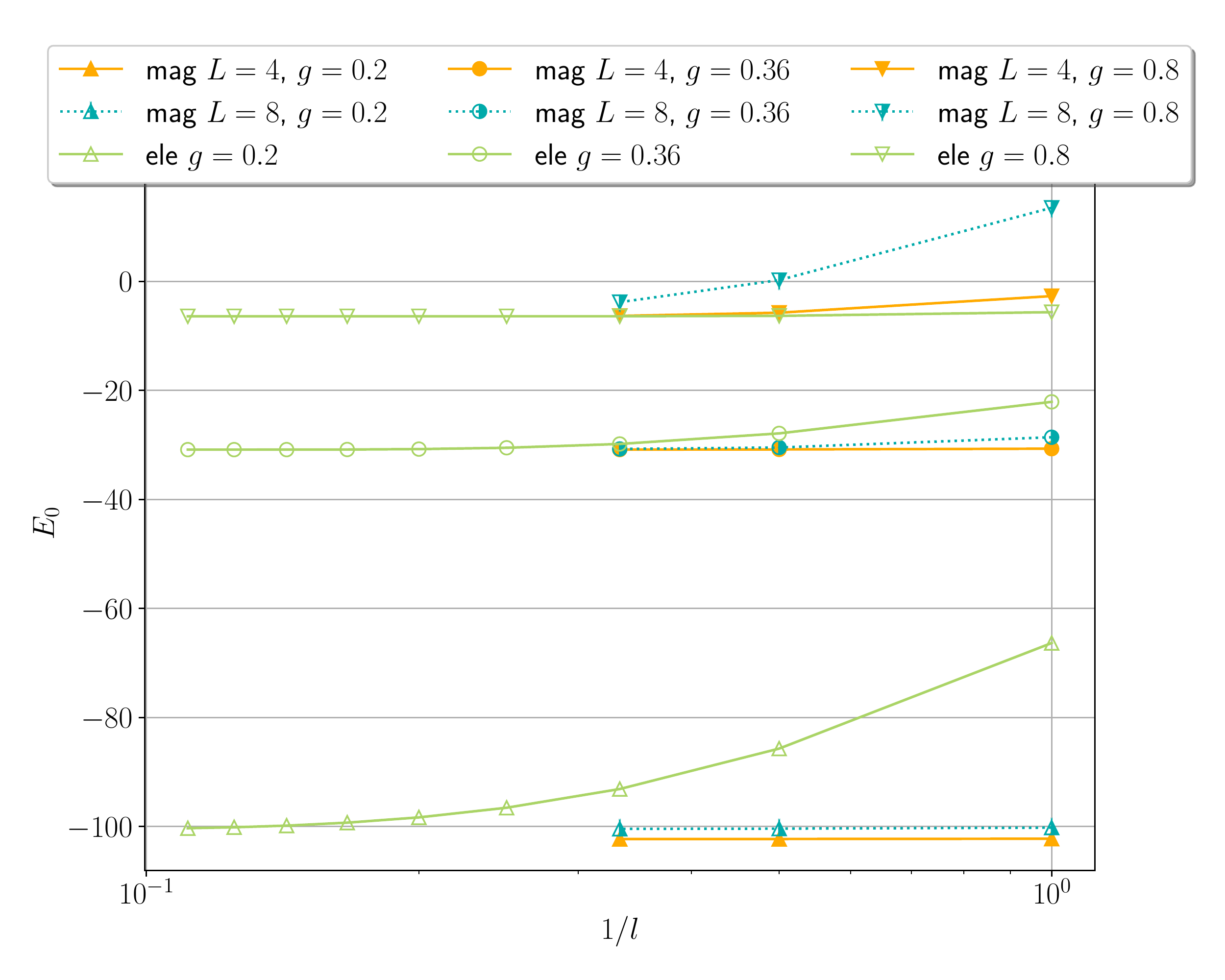}
    \caption{Exact diagonalization data for ground state energy as a function of $1/l$ for both electric and magnetic basis and selected values of $g$ (see legend).}%
    \label{fig:plot_match_ele_mag_QED_PBC_2x2_E_vs_1ol}
\end{figure}
\begin{figure}[ht]
    \centering
    \includegraphics[width=1\linewidth]{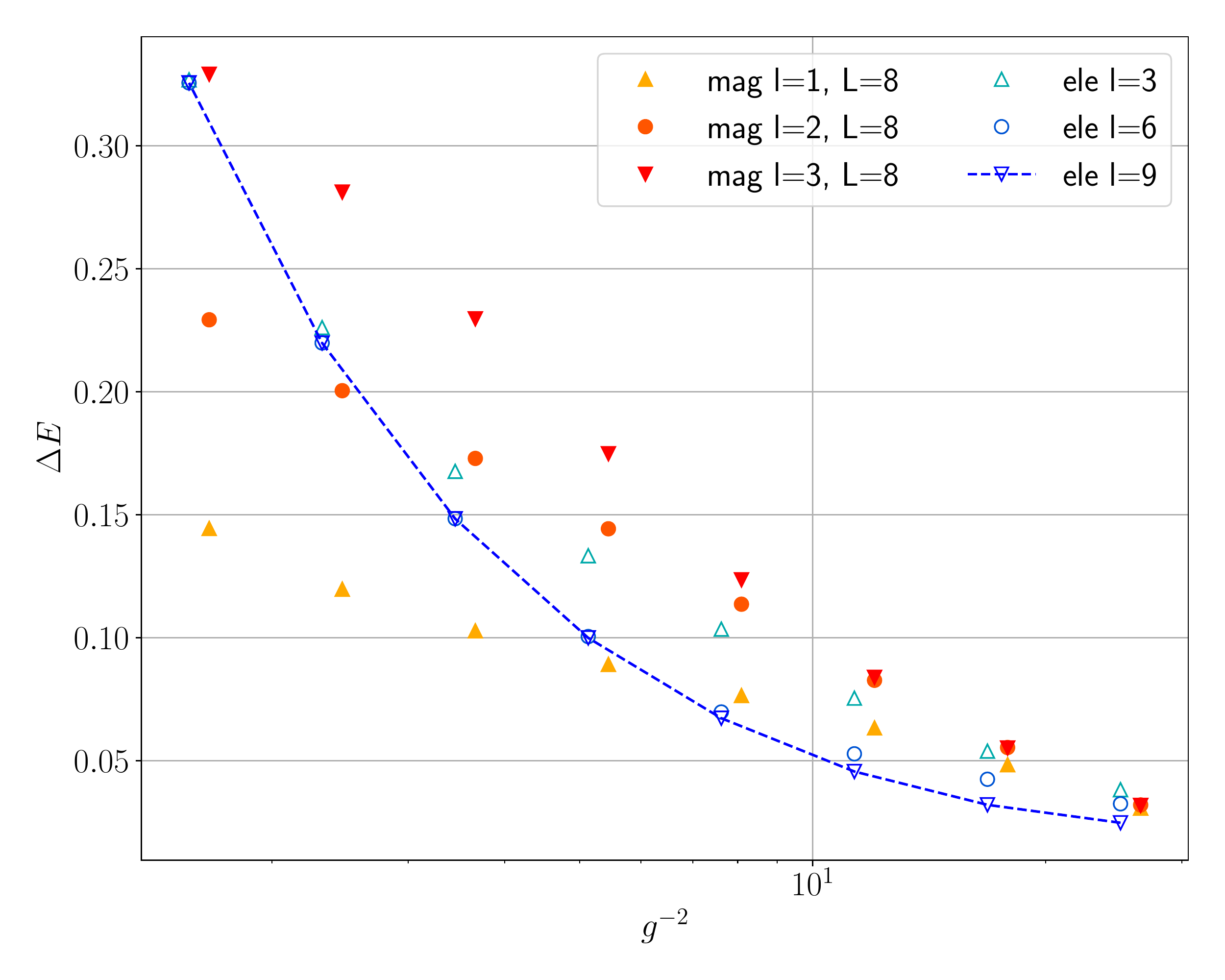}
    \caption{Exact diagonalization data for spectral gap in both electric and magnetic basis and selected values of discretization and truncation (see legend).}%
    \label{fig:plot_match_ele_mag_QED_PBC_2x2_dE}
\end{figure}
This behavior is apparent in Figure~\ref{fig:plot_match_ele_mag_QED_PBC_2x2_E_vs_1ol}, where we plot the estimates to the untruncated theory as functions
of the inverse truncation $1/l$ for some selected values of the coupling and for two values of $L$, i.e. 4 and 8. For the electric 
basis, the limit $l\to \infty$ is sufficient to 
discuss about convergence to the untruncated theory, 
while for the magnetic basis one has to consider 
a double limit where also the discretization level $L$ is taken into account.
However, analyzing the spectral gap
in the range $g\in [0.2,0.8]$, 
shown in Figure~\ref{fig:plot_match_ele_mag_QED_PBC_2x2_dE},
one sees a poorer convergence
in $l$ of the magnetic basis results with the 
electric ones, which for $l=3$ are already quite
close to the best estimate of the untruncated theory,
which is illustrated by the dashed line and is
represented by the electric basis result at $l=9$. For smaller values of $g$ we expect the
magnetic basis to be more convenient, 
but since the matching with MCMC results can 
be done for the couplings of $O(1)$ or larger,
we decided to consider mainly the
electric basis for the VQD results in the
following discussion.

\subsection{VQD results}
For the VQD results we adopted the 
NFT optimizer~\cite{PhysRevResearch.2.043158}
with the Qiskit's \textit{EfficientSU(2)}~\cite{Qiskit}
generic ansatz up to $5$ layers 
and with full entanglement gates (i.e., $CNOT$ gates between every pair of qubits) 
alternated to single-qubit rotation layers. Since we are testing the feasibility of the method, numerical results in this paper have been obtained using a simulator without noise and with an infinite number of shots.

\begin{figure}[ht]
    \centering
    \includegraphics[width=1.0\linewidth]{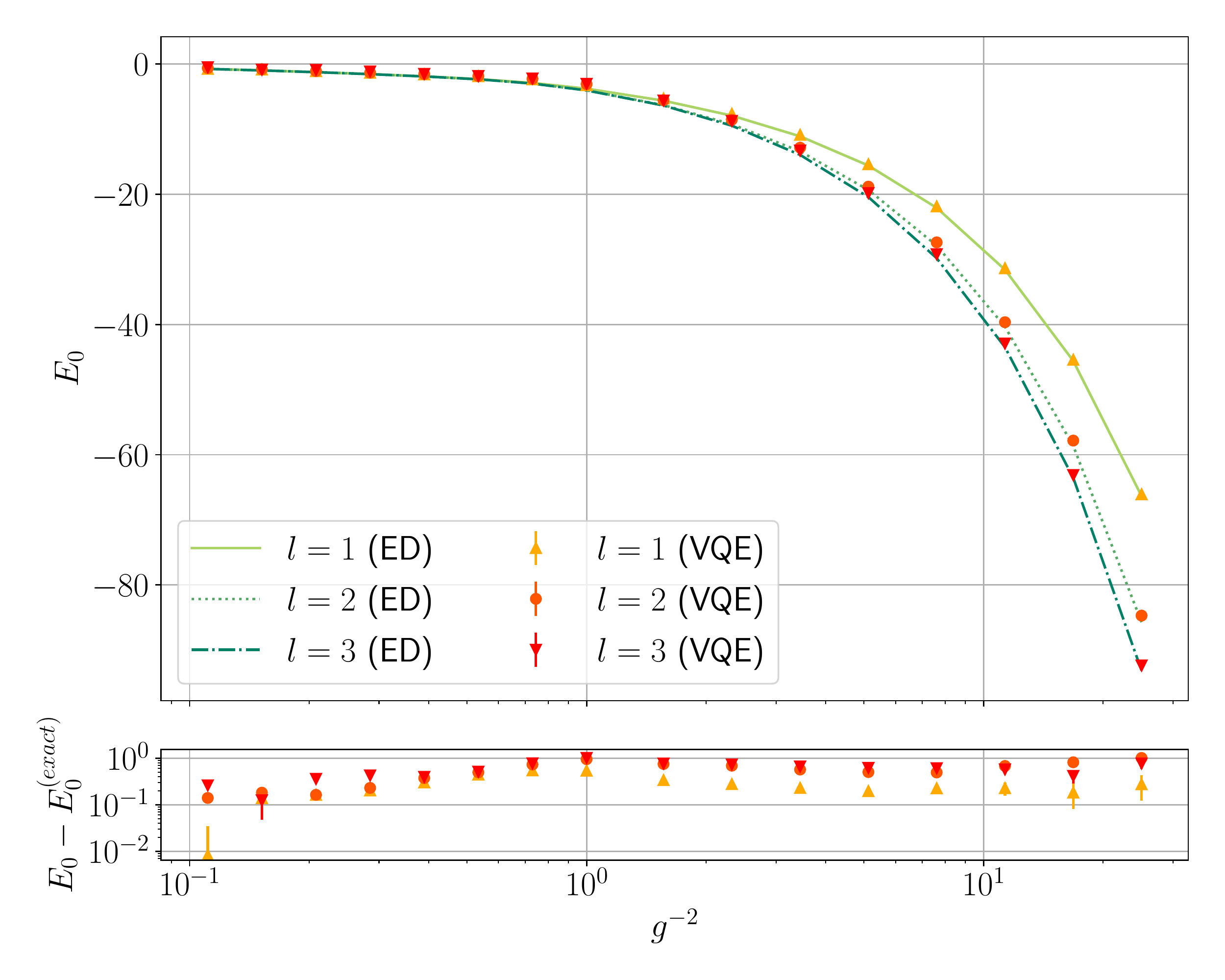}
    \caption{Best results for VQD ground state energy as a function of the coupling in the electric basis (dots) at some values of truncation level $l$ and exact diagonalization (lines). \textit{Bottom panel}: discrepancies with the exact values.}%
    \label{fig:plotvqe_egap_plaq_E0}
\end{figure}

\begin{figure}[ht]
    \centering
    \includegraphics[width=1.0\linewidth]{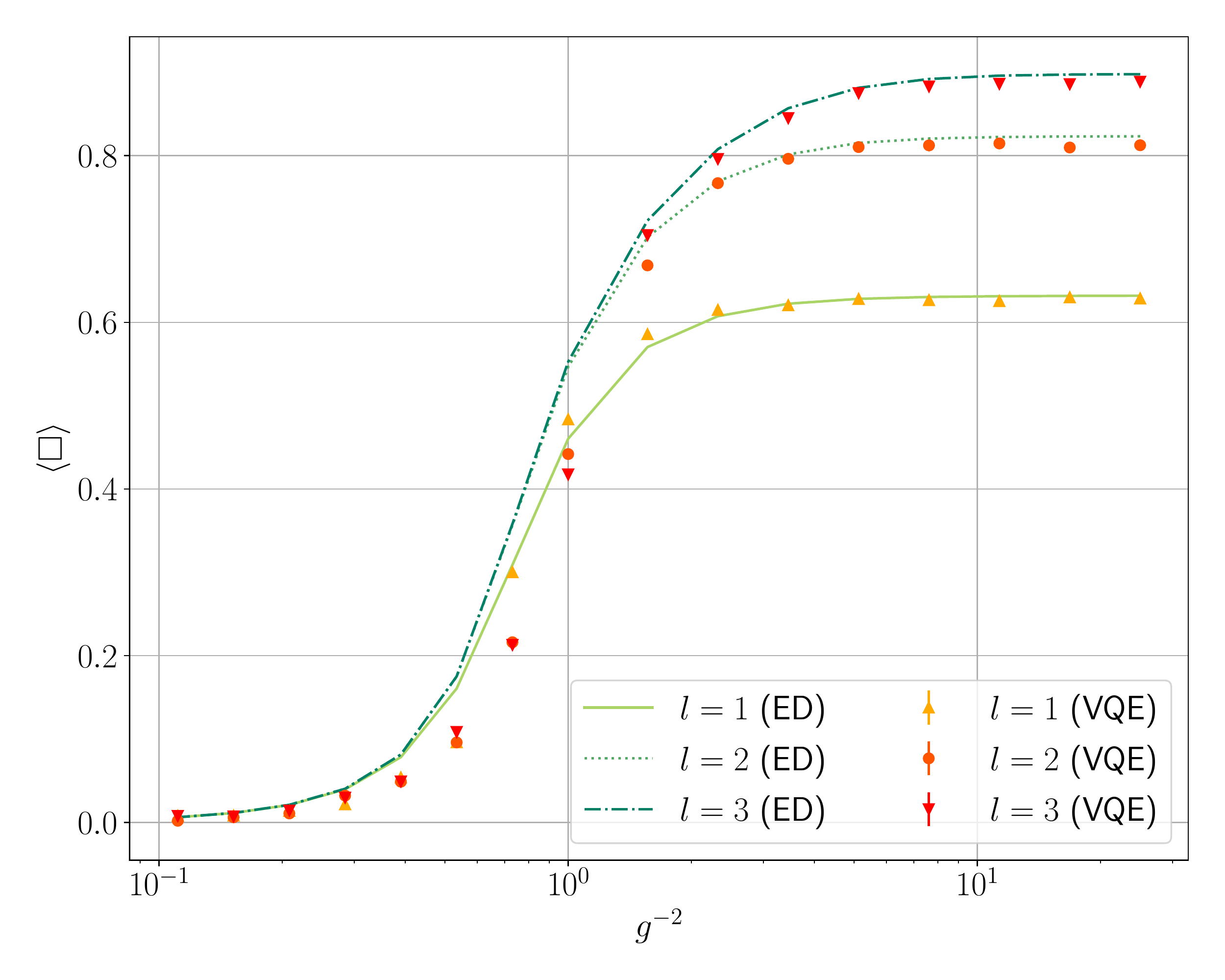}
    \caption{Plaquette measurements on the 
    ground state in Figure~\ref{fig:plotvqe_egap_plaq_E0} 
    as a function of the coupling $g$ ($m=0$) in the electric basis. VQE results (dots) and ED (lines).}%
    \label{fig:plotvqe_egap_plaq_Pl}
\end{figure}

\begin{figure}[ht]
    \centering
    \includegraphics[width=1.0\linewidth]{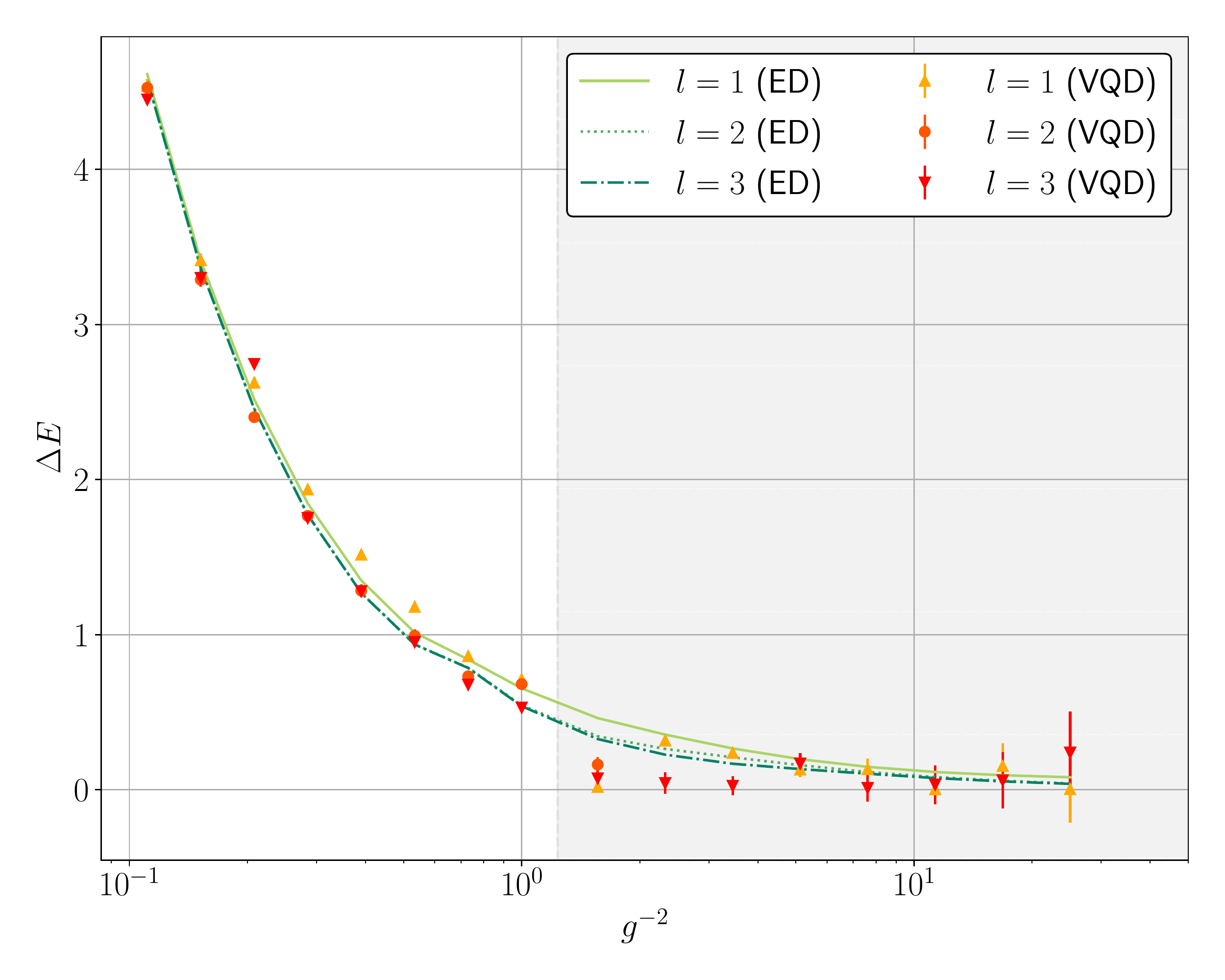}
    \caption{Best results for spectral gap as a function of the coupling $g$ ($m=0$) in the electric basis. VQD results (dots) and ED (lines); the shaded area corresponds to the region where we could not
    obtain results with enough precision to estimate
    the gap reliably.
    For all data points shown, the unphysical part of both the ground state and first excited does not exceed $0.5\%$.}%
    \label{fig:plotvqe_egap_plaq_DE}
\end{figure}

We first launched some runs to determine the best estimate to the ground state
and its energy. The results of the ground state energy and the plaquette as a function
of $g^{-2}$ in the range $[0.2,3]$ for  
$l=1,2,3$ in the electric basis are shown in Figure~\ref{fig:plotvqe_egap_plaq_E0} and~\ref{fig:plotvqe_egap_plaq_Pl}.
Then we used the best ground state obtained, for each $g$ and $l$, as a further penalty term in order to find the energy of the first excited state, as described in Sec.~\ref{subsec:VQAs}. From the exact result we have seen that the gap is quite small, thus we must be careful during the selection of the $\lambda$ factor and the number of iterations. As mentioned in Sec.~\ref{subsec:penalty},we consider a bisection method in order to choose the best value of $\lambda$. Starting from an initial value of $\lambda \sim O(1)$\footnote{The factor must be at least larger than the energy gap, as depicted in the pure gauge plots in Appendix~\ref{app:penaltygauge}.}, we compute the percentage of unphysical state and tune the suppression factor in terms of this percentage. In particular we choose a threshold of $99\%$ of physical component. 
Regarding the expectation value of the plaquette,
best results of the VQE (which minimize $E_0$) are shown in Figure~\ref{fig:plot_match_ele_mag_QED_PBC_2x2_Pl}.
The discrepancy between VQE and exact 
diagonalization results for $\expval{\Box}$ is 
larger especially in  the region close to $g=1$.
This may be explained by two competing factors.
On one hand, due to the closing of the gap 
(see Figure~\ref{fig:plot_match_ele_mag_QED_PBC_2x2_dE})
and the accumulation of quasi-degenerate levels in 
the spectrum for smaller values of $g$, 
the quality of the ground state obtained 
$\ket{\psi(\theta_0^*)}$
(and therefore also the estimate of the plaquette) 
is affected, since it is more likely to 
converge to a superposition
of quasi-degenerate states which cannot be
efficiently discriminated by the $VQE$ 
optimization process.
On the other hand, besides an irrelevant factor,
the plaquette observable coincides with the 
magnetic part of the Hamiltonian, which becomes
dominant in the regime of small $g$,
so that a decent estimate of the ground state 
energy $E_0$ 
(which is exactly what the VQE optimizes)
coincides with an acceptable estimate also for $\expval{\Box}$.
The ED and best VQD results of the spectral gap 
computed in the electric basis are shown in 
Figure~\ref{fig:plotvqe_egap_plaq_DE}
for truncation level up to $l=3$.
As mentioned previously,
the closing of the gap for $g < 1$,
corresponding to the shaded area on
the right of Figure~\ref{fig:plotvqe_egap_plaq_DE},
makes it difficult for the VQD 
to reach enough accuracy 
in the estimation of both the ground state 
and first excited levels.
In the region $g\geq 1$ the convergence
is more under control.
We conclude that a matching between MCMC data 
and VQD using the mass gap as observable
could be possible only in the region $g\geq 1$, 
because, due to physical features 
of the lower part of the spectrum, 
the accuracy required does not
allow to realistically extend the matching to
smaller values of $g$.
However, as we describe in the next section,
in order to extend the study of the running 
coupling to the weak coupling regime with $g < 1$,
we would use information from the static force
instead of the mass gap.
Since the ground state and first excited levels 
are well separated at values of the coupling 
larger than $g=1$, the first VQE optimization stage succeeds with higher 
probability in finding a good approximation $\ket{\psi(\theta^*_0)}$ to the ground state.
For values of the coupling smaller than $g=1$
the ground state and excited levels get closer and closer, making the optimization process harder,
since these states become quasi-degenerate and 
more iterations are required to distinguish the
optimal direction from the others in the lower spectrum. This can also be seen in the bottom panel of Figure~\ref{fig:plotvqe_egap_plaq_E0}, 
which shows the discrepancy between the VQE results and the exact values.

\section{Outlook: towards the running coupling}\label{sec:outlook}

As outlined in the previous sections, it is the main purpose of this work to establish the 
connection between quantum computations of short distance quantities and 
large volume Monte Carlo simulations which can provide us with the value of the 
lattice spacing. To this end, we 
have described here the setup for the quantum computation and showed that 
we can compute the mass gap in the relevant region of bare 
couplings where we expect to be able to match Monte Carlo simulations. 

As an important example of a short distance quantity we consider 
the running coupling. Despite the limited computational resources of the present quantum computer 
hardware, it is still
possible to provide definitions of the running coupling using  
the plaquette observable, the static potential or the static force at short distances.
In the following, we describe both approaches shortly, 
devoting a more
detailed discussion together with first numerical results to a future work. 

\subsection{Matching with MCMC data}\label{subsec:mcmc}
To set the scale, i.e., the physical values of the lattice spacing, 
we propose to take into consideration the spectral gap 
$\Delta E$ in the scalar sector and the static potential $V(r)$, which is defined as the lowest energy of a system of a static quark-antiquark pair at distance $r$.

As in QCD, from both lattice and experimental evaluations of $V(r)$, up to an unphysical offset, one can determine the analogue of the \emph{Sommer parameter} $r_0$ using the force~\cite{sommer1994new}:
\begin{equation}\label{eq:sommerpardef}
r_0^2 \frac{\partial V(r)}{\partial r}\Big\rvert_{r=r_0} = c,
\end{equation}
where $c$ is an suitably chosen constant which in QCD is conventionally fixed to $c^{QCD} = 1.65$
with a phenomenological value $r_0^{QCD} = 0.5$ fm.
In the
continuum, the expected form of the static potential in 2+1-dimensional QED
is the following~\cite{Loan_2003,Polyakov:1978vu,Gopfert:1981er}
\begin{equation}
    V(r) = V_0 + \alpha \log r + \sigma r,
    \label{eq:staticpotential}
\end{equation}
where the second term is a logarithmic Coulomb term 
while the third one represents 
the confinement term with string tension $\sigma$.  
On the lattice, can discretize the 
derivative in Eq.~\eqref{eq:sommerpardef} 
by finite differences 
$\frac{\partial V(r)}{\partial r}\simeq \frac{V(r_2)-V(r_1)}{r_2-r_1}$. 
Despite the current size 
limitations for lattices on quantum machines, 
it is sufficient to evaluate 
$V(r)$ using just $2$ distances,
which means $3$ lattice locations where 
the static charges are placed.
As we describe in more details in the following section, in general the QED results for
a single fermionic species depend on
both the bare gauge coupling $g$ and 
the bare fermionic mass $m$.
From the static potential computed on
a lattice one can estimate the static force at two
distances $r$ and as a function of the couplings $F(r;g,m)$. 
The force and the mass gap can be computed both, from MCMC simulations and in the Hamiltonian formalism. In order to match both approaches, the task is then to find matching values $(g^*,m^*)$ where the results for the mass gap the force agree. Such a matching is expected to be possible  in the intermediate coupling regime where, as we show in this paper with our Hamiltonian setup, 
we can reach sufficiently accurate results for the mass gap. A numerical investigation and matching 
of the static force will be discussed in a future work.

\subsection{Step scaling approach}\label{subsec:stepscaling}

One approach for computing the 
running coupling is to 
``scale step''~\cite{Luscher:1991wu,Luscher:1992an}, see~\cite{Luscher:1998pe} for 
an introduction to the Schr\"odinger functional scheme 
for the step scaling procedure.

As an example, we consider the coupling defined from the static potential.
Let us take, for illustration purposes
a 
Coulomb potential\footnote{The here discusses example can be straightforwardly extended to the case of 2+1-dimensional QED with the potential given by Eq.~\eqref{eq:staticpotential}.} $V(r)=\alpha(r)/r$ and a pure gauge theory with 
bare coupling $g$. Then, a running coupling at two scales 
$r_1$ and $r_2$ can be defined by $\alpha_{\mathrm ren}(r_1)=r_1 V(r_1)$ 
and $\alpha_{\mathrm ren}(r_2)=r_2 V(r_2)$. 
Let us now assume that we have computed $\alpha_{\mathrm ren}(r_1^{i=0},g_0)$ and 
$\alpha_{\mathrm ren}(r_2^{i=0},g_0)$ at a fixed value of the bare coupling $g^2=g_0^2$ at a 
step $i=0$.
The two distances can be related by a scale factor $s$, e.g. $s=2$ such that   
$r_2^{i=0}= s r_1^{i=0}$. 
In the next step, $i=1$, the bare coupling $g^2$ is tuned 
in such a way that 
$\alpha_{\mathrm ren}(r_1^{i=1},g)=\alpha_{\mathrm ren}(r_2^{i=0},g_0)$ which would 
provide a value $g_1^2$ where the renormalized couplings agree and hence
the scales match. Applying the 
same scale factor $s$ at the found bare coupling $g_1^2$ one arrives at the renormalized coupling 
at the scale $r_2^{i=1}=s r_1^{i=1}=2 s r_1^{i=0}$. Thus we get sequence of renormalized 
couplings $\alpha_{\mathrm ren}(r_1^{i=0},g_0)$, $\alpha_{\mathrm ren}(2 s r_2^{i=1},g_1)$. 

This procedure can be repeated $N$ times such that we obtain the scale 
dependence of the coupling where the scale changes by the factor $s$ in each step 
arriving thus at a renormalized coupling $\alpha_{\mathrm ren}(N s r_1^{i=0},g_\mathrm{match})$
with $g_\mathrm{match}$ chosen large enough that contact with Monte Carlo simulations
can be made. Note that up to this point we have only worked with quantities 
in lattice units thus not knowing the physical values of the scale. 

It is exactly at this point where the matching of the quantum and 
the classical Monte Carlo computations
comes into play. The Monte Carlo simulation can be performed in large
volumes at intermediate values of the coupling and through the 
strategy described below a value 
of the lattice spacing can be determined. In this way, the final 
scale can be converted to physical units, i.e. 
$r_\mathrm{phys}=a N s r_1^{i=0}$, given thus the renormalized 
coupling $\alpha_{\mathrm ren}(r_{\mathrm{phys}},g_{\mathrm{match}})$. The 
sequence of couplings obtained as outlined  above can now be inverted   
by subsequently changing the scale by a   
factor $s$, i.e. $r_\mathrm{phys}^{i=N}\rightarrow r_\mathrm{phys}^{i=N-1} = r_\mathrm{phys}^{i=N}/s$ 
$\rightarrow r_\mathrm{phys}^{i=N-2} = r_\mathrm{phys}^{i=N-1}/s$, ..., $\rightarrow r_\mathrm{phys}^{i=0}$
with the corresponding changes of the coupling.
In this way, the renormalized coupling is obtained as a function of the 
physical scale and, by making contact with perturbation theory, 
eventually also the important $\Lambda$ parameter, which provides 
the scale where non-perturbative physics sets in, can be determined in 
physical units. 

The procedure explained here can, in principle, also be used to
disentangle lattice effects from the real running of the coupling by 
taking the continuum limit. However, this would require 
large lattices and, as mentioned already, with present
quantum computing hardware resources this is not feasible 
and would require future quantum computers with more and improved 
--ideally error corrected-- qubits.
Nevertheless, the just described procedure can be implemented 
on already existing quantum devices allowing thus to 
go to the deep perturbative regime and to make contact 
to low order perturbation theory.

\subsection{Boosted coupling approach and scale setting}\label{subsec:boosted}

\begin{figure}[ht]
    \centering
    \includegraphics[width=1\linewidth]{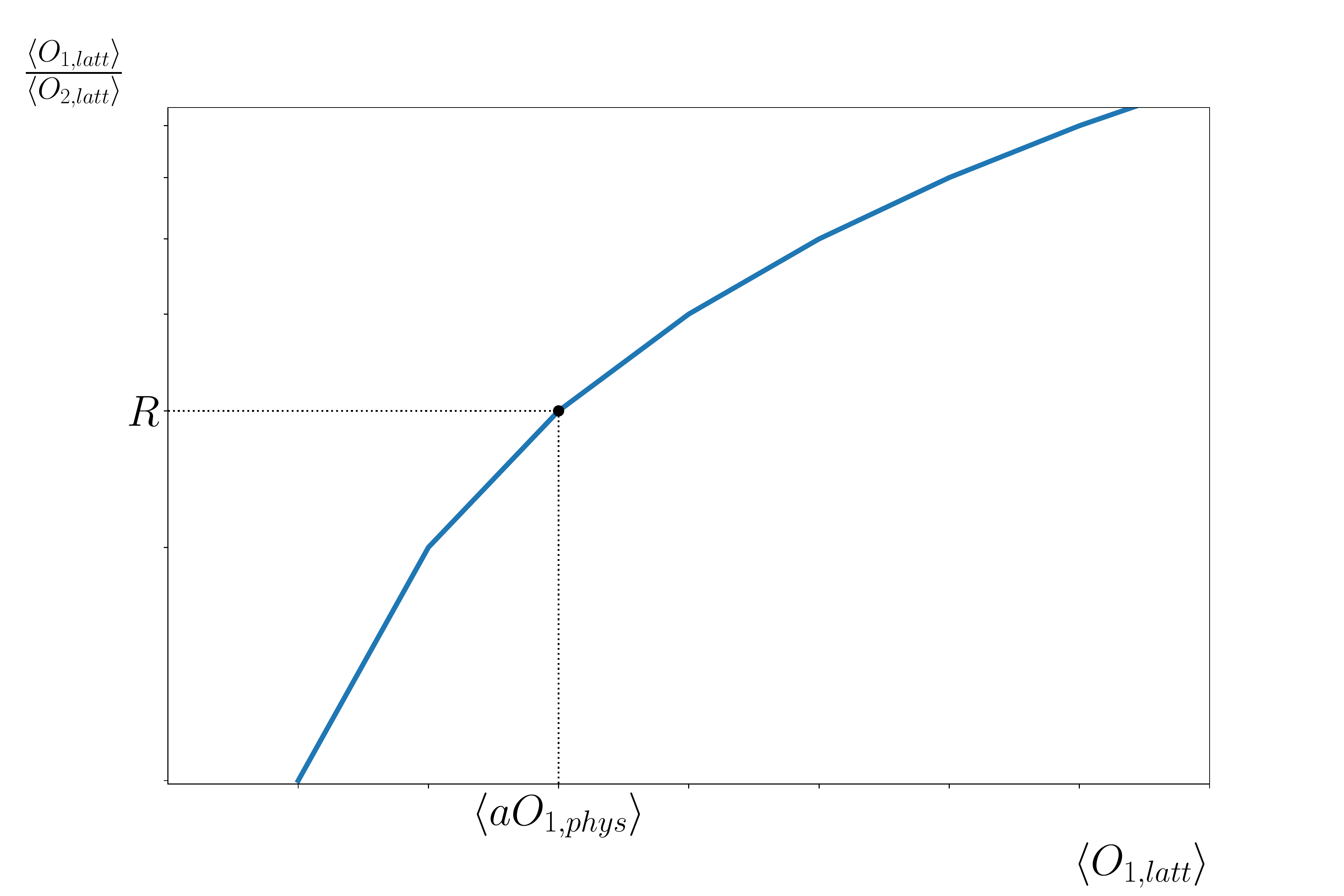}
    \caption{Illustration of principle way to determine a value
    of the lattice spacing, see discussion in main text.}%
    \label{fig:ratio}
\end{figure}

An alternative way 
to determine the running coupling is to employ  
the boosted coupling defined by 
\begin{equation}
    g^2_{\rm boosted} = g^2/\langle \Box\rangle\; ,
    \label{eq:boosted} 
\end{equation}
where $\expval{\Box}$ denotes the expectation value of the plaquette operator and $g^2$ is the bare gauge coupling.
Following the strategy given in ~\cite{BOOTH2001229} the scale dependence of the coupling 
can then be determined employing perturbation theory to a given order. 
An essential element in this procedure is the determination of 
the renormalization scale in physical units for which the lattice spacing needs 
to be calculated. The general setup for such a calculation is illustrated in Fig.~\ref{fig:ratio}
where $\expval{O_1}$ and $\expval{O_2}$ denote 
expectation values of two observables\footnote{The ratio $R$ ought be dimensionless, meaning that 
$O_{1,\mathrm{phys}}$ and $O_{2,\mathrm{phys}}$ have the same mass dimension. In Fig.~\ref{fig:ratio} we assume for simplicity a mass dimension of one in order to relate the lattice value $\langle O_1^\mathrm{latt}\rangle$ to the physical one, $\langle O_1^\mathrm{phys}\rangle$, and hence extract the lattice spacing.} which can, in principle, 
be extracted from experiments. Thus the physical value of the ratio 
$R=\expval{O_1}/\expval{O_2}$ is known 
(indicated by a "$\bullet$" in Fig.~\ref{fig:ratio}). 
One can then tune the parameters of the theory such that at a 
certain value of $\expval{O_{1,\mathrm{latt}}}$, $R$ is reproduced. 
At this value of $\expval{O_{1,\mathrm{latt}}}$ 
the lattice spacing can be determined by the relation $aO_{1,\mathrm{phys}}=O_{1,\mathrm{latt}}$.

Examples of $O_1$ and $O_2$ are particle masses (mass gaps) or the static force
at a given physical distance. However, many more choices are 
possible and actually used in large scale  lattice simulations. 
As we show in this paper, mass gaps 
become increasingly difficult to determine when the coupling is decreased towards 
the continuum limit and hence we, unfortunately, consider this not to be an option when we want 
to work in the perturbative regime. 

As an alternative we consider the static force at small distances. 
In this case, we can define $\langle O_{j,\mathrm{phys}}\rangle$ as $F_{\mathrm{phys}}(r_{j,\mathrm{phys}})$. A difficulty 
for using this strategy is that on the lattice the physical distance 
$r_{\mathrm{phys}}=a N$, with $N$ the number of lattice points, needs to be kept 
constant. This can be achieved, e.g., by demanding that $r^2F(r) = \mathrm{constant}$. As a consequence, 
when going to smaller values of the lattice spacing, the number of lattice 
points needs to be increased correspondingly. We therefore consider the 
possibility of using the static force for setting the scale as a conceptually 
clean way which can, however, only be employed when significantly larger future 
quantum computing resources are available. 
Hence, we consider the here described strategy to compute the running 
coupling as a nice but only future perspective.

\section{Conclusions}\label{sec:conlusions}

In this work we have provided a first step to combine 
classical Markov Chain Monte Carlo simulations with quantum computations. 
While MCMC calculations can be performed on very large lattices 
nowadays, they are limited to reach small values of the lattice spacing. 
On the other hand, quantum computations which uses the Hamiltonian formulation 
can, at least in principle, be used at arbitrary lattice spacings. However, 
presently, quantum computations are limited to small systems given the 
present generation of noisy quantum computers. They are therefore 
restricted to small distance quantities such as the running coupling 
or scale dependent renomalization functions.

It would therefore be ideal to combine both approaches and perform quantum computations at small values of the lattice spacing and large volume MCMC simulations to provide the physical scale. To this end, 
in this paper we have

\begin{itemize}
\item developed a resource efficient encoding for 2+1 dimensional QED, which allows to simulate the 
model eventually on already now available or forthcoming near term quantum computer with 
more and better qubits;
\item demonstrated that the introduction of suitable suppression terms can be used to force the final state of the optimization of a generic ansatz to have a small overlap with the unphysical Hilbert space; 
\item obtained results for the ground state energy in broad range of coupling which in turns allows 
us to compute the plaquette expectation value, or the static force at small distances which can be related to the running coupling;
\item shown that we can obtain accurate enough results for the energy gap in the large and intermediate coupling regime, which provides an important first step to eventually to make the desired contact to MC simulations.
\end{itemize}

We also want to remark that the 
the 
setup of 2+1-dimensional QED developed here will serve as the basis for extensions such as adding topological terms, chemical potential or real time simulations, directions, we want to follow in the future.   

In this work we set the stage for an 
application of quantum computing (using VQE in particular) 
to study lattice gauge theories 
in a non-perturbative fashion. The present work addresses ground state properties as well as the mass gap of 2+1-dimensional QED with the aim to reach small values of the lattice spacing without running into problems with autocorrelations. 
However the main advantage of this paradigm comes from 
the possibility of being applied to study
systems with numerical sign problems, 
which poses a challenge to standard MCMC methods.
In future works, given the here developed setup, we will
explore the addition of a
chemical potential and topological 
$\theta$ term to the $(2+1)$-dimensional QED Hamiltonian, being thus able to go far beyond traditional MCMC simulations -- at least when quantum hardware will be available to simulate large lattices.  

\begin{acknowledgments}
A.C. is supported in part by the Helmholtz
Association - “Innopool Project Variational Quantum
Computer Simulations (VQCS)”.
This work is supported with funds from the Ministry of Science, Research and Culture of the State of Brandenburg within the Centre for Quantum Technologies and Applications (CQTA).
\end{acknowledgments}

\bibliography{bibl} 
\bibliographystyle{ieeetr}

\appendix

\section{Structured ansatz for incomplete Gray encoding}\label{app:gray_ansatz}
Here we discuss how a parameterized circuit should be adapted in order to allow transitions only
between encoded states.

For the cases where states are encoded using the computational basis, 
the ans\"atze can be built by identifying the states to be excluded.
\begin{figure}[ht]
    \centering
    \begin{center}
    \subfloat{\includegraphics[width=0.9\linewidth, valign=c]{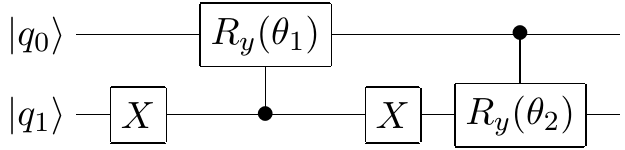}}
    \end{center}
\caption{Parametric circuit for cycling between encoded states with the Gray encoding at truncation $l=1$.}\label{fig:pcirc_gray1}
\end{figure}
\begin{figure}[ht]
    \centering
    \begin{center}
    \subfloat{\includegraphics[width=1.0\linewidth, valign=c]{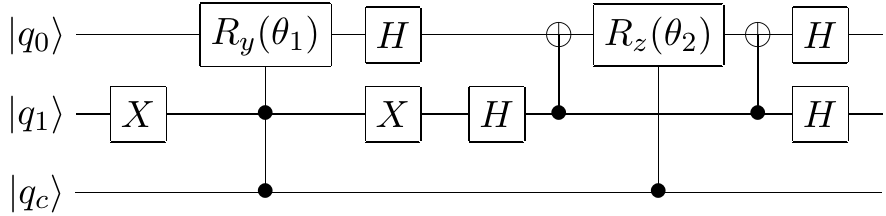}}
    \end{center}
\caption{Controlled parametric circuit for cycling between encoded states with the Gray encoding at truncation $l=1$.}\label{fig:pcirc_gray1_control}
\end{figure}
For example, in the case of Gray encoding with $l=1$, shown in Table~\ref{tab:gray_enc1},
one can cycle between the three encoded states for each link $\{\ket{00},\ket{01},\ket{11}\}$,
using the parametric circuit shown in Figure~\ref{fig:pcirc_gray1}, 
where the two rotations parameterized by $\theta_1$ and $\theta_2$ drive the transitions 
$\ket{00}\leftrightarrow \ket{01}$ and $\ket{01}\leftrightarrow \ket{11}$ respectively,
while the parametric circuit shown in Figure~\ref{fig:pcirc_gray1_control} entangles 
the gauge register ($q_1,q_0$) with a matter component ($q_c$),
by rotations which make for the transitions $\ket{00}\leftrightarrow \ket{01}$ 
and $\ket{00}\leftrightarrow \ket{11}$ (the rotation in $\theta_2$ is actually a 
$CR_{XX}(\theta_2)$).

\section{Antipodal state as physical in the magnetic basis}\label{app:fullcompact_antipodalmagnetic}
As described in Sec.~\ref{subsubsec:truncation}, the discretization adopted in~\cite{Haase2021resourceefficient} to 
write the Hamiltonian in the magnetic basis 
involves an uniform grid of $2L+1$ sites
on the group $U(1)$, where the identity 
is represented by $\ket{0}_{\text{phys.}}$.
From these states, only the $2l+1$ around 
the identity are selected as physical ones.
However, as already mentioned 
in Sec.~\ref{subsec:encoding},
being the number of physical states 
not a power of two makes the optimization problem 
harder, either at the Hamiltonian or at the ansatz level.
Including states non-symmetrically with respect to the 
identity (or origin in the electric basis) would break the charge conjugation symmetry of the
Hamiltonian, therefore introducing another problem.

A straightforward solution, 
although working only in the magnetic case,
would be to change the discretization procedure: 
instead of using $2L+1$ sites (i.e., the phases $\phi_j\equiv \frac{2\pi j}{2L+1}$), one can 
use a $2L+2$ grid, in such a way that the site 
antipodal to the identity is included in the discretization. Then, at the truncation stage,
besides the $2l+1$ states around the identity
we can include also the remaining $2^{\lceil \log_2(2l+1) \rceil} -2l-1$ 
as states around the antipodal to the identity. 
In this way, all the states used 
are encoded as physical states belonging 
to the discretization grid and one can use a generic
ansatz without having to penalize unphysical states
for the gauge variables.

Another advantage of this discretization
into $2L+2$ states is that 
it is possible to increase the truncation 
level from $L$ to $L^\prime=2L+1$ 
by keeping the previous grid 
and inserting sites at its midpoints.
Unlike the $2L+1$ grid, 
which cannot be simply related to finer
grids at larger $L$,
this would allow for a better control on
the extrapolation to infinite 
discretization.

\section{Penalty term method }\label{app:penalty}
We first discuss the introduction of a penalty term in the pure gauge system with periodic system and then we mention some information about the case with fermions.

\subsection{Test in pure gauge theory}\label{app:penaltygauge}
\begin{figure}[ht]
    \centerline{\includegraphics[width=1.2\linewidth]{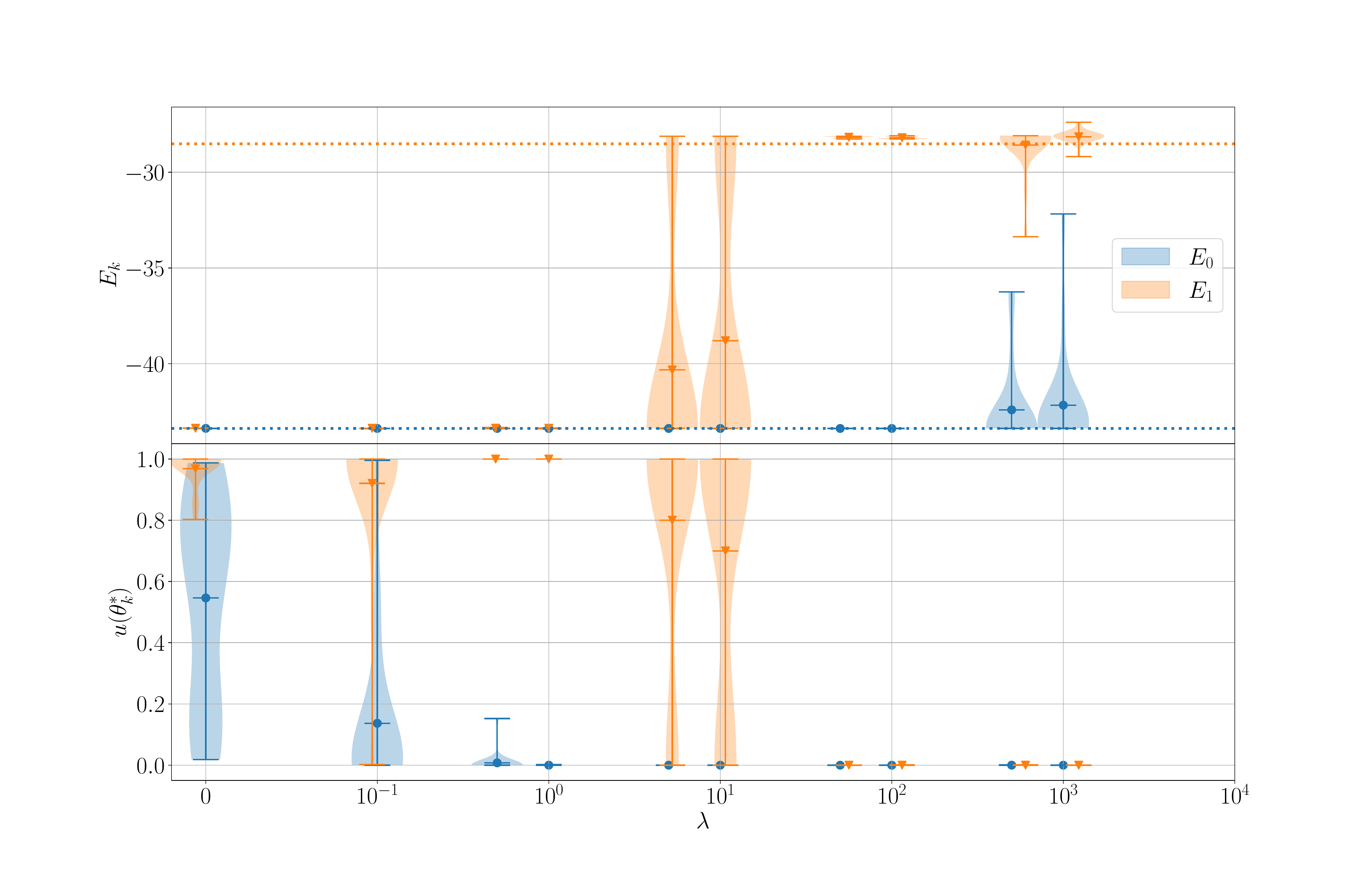}}
    \caption{Suppression factor method for a given value of the coupling. Energy eigenvalues from variational approach (optimization with 200 iterations and average of 20 runs) and exact diagonalization (dotted lines in \textit{upper panel}) and amount of unphysical states in the VQD solution (\textit{lower panel}).} 
    \label{fig:cutoff_plt200}
\end{figure}

In the upper panel of Figure~\ref{fig:cutoff_plt200}
it is depicted the trend of the energy eigenvalues ($E_0$ in \textit{blue} and $E_1$ in \textit{orange}) for a certain range of $\lambda$. 
When the suppression coefficient $\lambda$ is zero the first excited state is almost degenerate to the ground level. 
However, one can see in the corresponding value of unphysical states in the lower panel, that the result does not represent a physical solution.
\begin{figure}[H]
    \centerline{\includegraphics[width=1.2\linewidth]{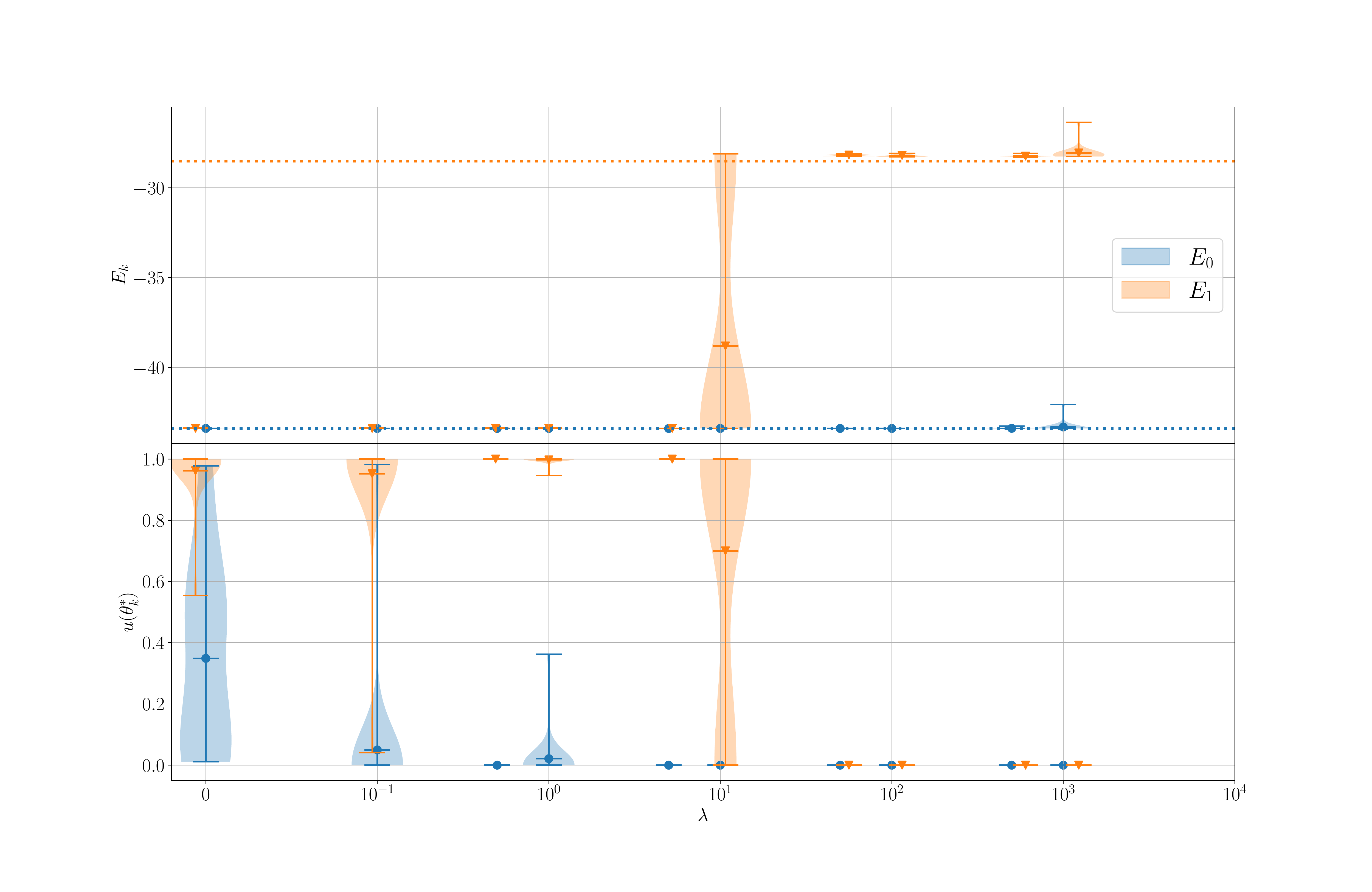}}
    \caption{Suppression factor method for a given value of the coupling. With 800 iterations we increase the possibilities to obtain accurate results when performing a variational method.} 
    \label{fig:cutoff_plt800}
\end{figure}
\begin{figure}[H]
    \centerline{\includegraphics[width=1.2\linewidth]{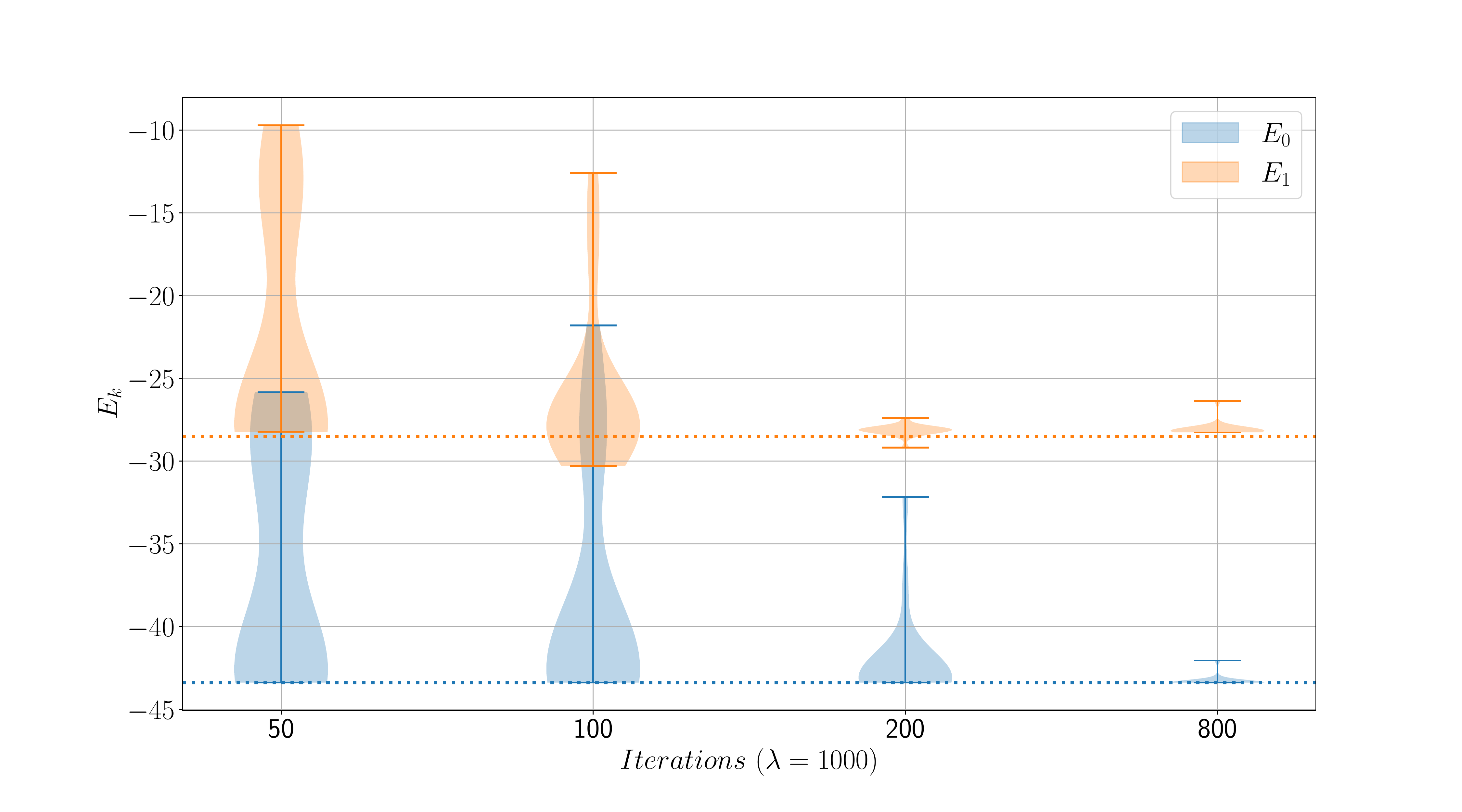}}
    \caption{Trend of VQD results with fixed $\lambda=1000$ and different values of iterations, (50, 100, 200 and 800).} 
    \label{fig:cutoff_pltl1000}
\end{figure}
As $\lambda$ increases, the percentage of unphysical states decreases, until we can reach an accurate solution after the unphysical part of $E_0$ and $E_1$ is suppressed, ($\lambda\sim O(10)$).
However, it is important to not have a large $\lambda$ when performing the variational method. This because the algorithm cannot reach a low, acceptable value for the energies. In the rightmost part of the plot the results deviate from the exact solution, even if we can avoid unphysical states, the fidelity decreases\footnote{In some runs, it may happen that the values found for the first excited is lower than the one of the ground state. In this case one can exchange the two states and values.}.
If we consider more iterations for the optimization process, the possibilities for the optimizer to converge increases even if $\lambda$ is large, as depicted in Figure~\ref{fig:cutoff_plt800} and Figure~\ref{fig:cutoff_pltl1000}.
\begin{table}[h]
    \caption{Energy eigenvalues $E_k$, in the coupling range $g\geq 1$ for electric basis with truncation $l=1$. Also the percentage of unphysical states, $u(\theta^*_k)$ and number of layers $n_{\text{lay.}}$ in the quantum circuit are shown.}
    \centering
    \begin{tabular}{c|c|c|c|c|c|c|c}
        $g$ & $n_{\text{lay.}}$ & $E_0$ & $E_0^{(\text{exact})}$ & $u(\theta^*_0)$ & $E_1$ & $E_1^{(\text{exact})}$ & $u(\theta^*_1)$ \\ \hline
        1.0     & 3 & -3.173(24) & -3.799 & 0.0   & -2.461(23) & -3.145 & 0.39 \\
        1.16993 & 4 & -2.272(24) & -2.915 & 0.195 & -1.410(18) & -2.075 & 0.293 \\
        1.36874 & 4 & -1.89(2)   & -2.328 & 0.391 & -0.715(18) & -1.315 & 0.195 \\
        1.60133 & 5 & -1.59(2)   & -1.911 & 0.098 & -0.073(18) & -0.561 & 0.0 \\
        1.87344 & 4 & -1.324(21) & -1.568 & 0.098 & 0.75(2)    &  0.279 & 0.195 \\
        2.1918  & 5 & -1.085(23) & -1.265 & 0.0   & 1.539(27)  &  1.251 & 0.098 \\
        2.56425 & 5 & -0.847(25) & -0.998 & 0.0   & 2.567(36)  &  2.412 & 0.098 \\
        3.0     & 5 & -0.698(27) & -0.769 & 0.0   & 3.841(37)  &  3.844 & 0.0
    \end{tabular}\label{tab:fermresult}
\end{table}

\subsection{Penalty term for the fermionic system}\label{app:penaltyferm}
For the fermionic case we applied a specific approach and considered only a small range for the suppression parameter. In Table~\ref{tab:fermresult} we reported the best results for the ground state energy and first excited state for the truncation level $l=1$ and in the coupling range $g\in [1,3]$. 
The main criterion was to set a threshold for the percentage of unphysical states. No explicit dependence on the number of layers or small variations of $\lambda$ was found;
this might be due to the variability of the optimization process.

\end{document}